\documentclass[12pt,a4paper]{article}
\usepackage{ifpdf}
\ifpdf
\pdfpageattr {/Group << /S /Transparency /I true /CS /DeviceRGB>>}
\fi
\usepackage{jheppub}
\usepackage{amsmath}
\usepackage{caption}
\usepackage{subcaption}

\def\bZ{\mathbb{Z}}
\def\bR{\mathbb{R}}

\newcommand{\beq}{\begin{equation}}
\newcommand{\eeq}{\end{equation}}
\newcommand{\bea}{\begin{eqnarray}}
\newcommand{\eea}{\end{eqnarray}}

\newcommand{\RR}{{\mathbb R}}

\def\to{\rightarrow}

\newcommand{\fsl}{\mathfrak{sl}}
\newcommand{\SL}{\mathrm{SL}}
\newcommand{\ISO}{\mathrm{ISO}}
\newcommand{\U}{\mathrm{U}}

\begin{document}

\title{Strings vs Spins on the Null Orbifold}
\author[a]{K. Surya Kiran,} \ 
\author[a]{Chethan Krishnan,} \
\emailAdd{ksuryakn@gmail.com}
\emailAdd{chethan.krishnan@gmail.com}
\affiliation[a]{Center for High Energy Physics, \\
  Indian Institute of Science, \\ 
  Bangalore - 560012, \ \ India}
\author[a,b,c]{Ayush Saurabh,}
\affiliation[b]{International Center for Theoretical Sciences, \\
  Indian Institute of Science Campus, \\ 
  Bangalore - 560012, \ \ India}
\affiliation[c]{School of M. A. C. E., The University of Manchester,\\ 
  Manchester - M13 9PL, \ \ United Kingdom}
\emailAdd{ayushsaurabh@hotmail.com}
\author[d]{Joan Sim\'on}
\affiliation[d]{School of Mathematics and Maxwell Institute for Mathematical Sciences,\\
University of Edinburgh, King's Buildings, \\
Edinburgh EH9 3JZ, UK}
\emailAdd{j.simon@ed.ac.uk}
\keywords{Big-Bang, Higher Spin Theories, String Scattering Amplitudes}
\abstract{We study the null orbifold singularity in 2+1 d flat space higher spin theory as well as string theory. Using the Chern-Simons formulation of 2+1 d Einstein gravity, we first observe that despite the singular nature of this geometry, the eigenvalues of its Chern-Simons holonomy are trivial. Next, we construct a resolution of the singularity in higher spin theory: a Kundt spacetime with vanishing scalar curvature invariants. We also point out that the UV divergences previously observed in the 2-to-2 tachyon tree level string amplitude on the null orbifold do not arise in the $\alpha^\prime\to \infty$ limit. We find all the divergences of the amplitude and demonstrate that the ones remaining in the tensionless limit are physical IR-type divergences. We conclude with a discussion on the meaning and limitations of higher spin (cosmological) singularity resolution and its potential connection to string theory.
}

\setcounter{tocdepth}{2}
\maketitle

\section{Introduction and Summary}\label{introduction}

The understanding of cosmological singularities is one of the most important questions in quantum gravity. Little progress has been achieved so far in string theory regarding this issue. The simplest toy models based on Lorentzian orbifolds of flat space\footnote{Reviews can be found in \cite{Cornalba:2003kd, Durin:2005ix, Craps:2006yb, Berkooz:2007nm}.} give rise to UV divergent string scattering amplitudes that were interpreted as a breakdown of string perturbation theory due to uncontrolled backreaction at the singularity \cite{LMS, Lawrence:2002aj, HoroPolch, Ben}. 

Recently, the possibility of resolving toy versions of the big-bang singularity 
was studied in \cite{Shubho1, Shubho2, Avinash} by embedding 2+1 de Sitter or flat space quotient singularities in higher spin theories\footnote{In 2+1 dimensions higher spin theories have a formulation in terms of Chern-Simons theories \cite{Campoleoni}. The flat space versions were constructed in \cite{Arjun, Troncoso}.}. The underlying motivation for this is that higher spin theories are expected to capture aspects of the tensionless limit of string theory. 
So one might wonder whether by working at small string coupling but including the infinite number of massless string modes that arise in this limit, the behaviour of the singularity can be tamed in some way\footnote{This is plausible because the infinite number of massless states signal the presence of enhanced gauge symmetries in spacetime, that subsume the usual diffeomorphism invariance of gravity.}. Evidence for this in the affirmative has been provided in two different regimes for the Milne singularity, which is a boost orbifold of flat pace: in the target space description using higher spin theory \cite{Shubho1, Shubho2, Avinash}, and in the low tension limit of string scattering amplitudes \cite{BenAyush}.

In an effective target space description, one expects the emergence of a higher spin theory with an enhanced gauge symmetry in the tensionless string limit. Various arguments have been proposed to make this correspondence precise in AdS \cite{Vasiliev, Sundborg, Witten,Chang:2012kt}. Even though there is no such concrete understanding in the flat space case,  one expects that the 2+1 dimensional higher spin theory presented in \cite{Arjun, Troncoso} should capture {\em some} aspects of flat space string theory in the tensionless limit. One then looks for classical configurations having the same holonomy as the singular cosmology, so that they are physically equivalent, but in which one can find gauges where the metric is non-singular. This is in the same spirit as higher spin black hole singularity resolutions in \cite{Gutperle, Maloney}. 

In string perturbation theory, one can scan for divergences in the known Lorentzian orbifolds and study whether the behaviour of these divergences gets tamed in the $\alpha^\prime \to \infty$ limit. In particular, one can study whether any surviving divergences in this limit are physically acceptable infrared divergences including those arising from the tower of intermediate string states going on-shell.

This programme was satisfactorily carried out for the boost orbifold (Milne universe). The existence of higher spin classical resolutions was reported in \cite{Shubho1, Shubho2} and the absence of un-physical UV divergences in the tensionless string limit was discussed in \cite{BenAyush}. In this work, we study the same issues for the null orbifold, the unique 3d supersymmetric Lorentzian abelian orbifold having no closed timelike curves \cite{JoseJoan}, and reach similar conclusions. Our main results are :
\begin{itemize}
\item The Chern-Simons (C-S) holonomy of the null orbifold is trivial. This is an explicit proof of principle that the pathology of a singularity is not necessarily reflected in its holonomy. 
\item A possible higher spin resolution of the null orbifold is a Kundt space-time supported by higher spin fields, which has no non-vanishing curvature scalars.
\item We point out that the previously identified UV divergences in the 4-pt string scattering amplitude \cite{LMS} are no longer there in the large-$\alpha^\prime$ limit. We do an exhaustive scan of the divergences and show that the ones that remain in the low tension regime are IR divergences
that are expected on physical grounds.
\end{itemize}

It is worth stressing that despite the rather different structure of the string scattering amplitudes, we find a detailed correspondence between the structure of the divergences in Milne \cite{BenAyush} and here. It would be interesting to have a universal understanding of the origin of these divergences in terms of statements about strings in the covering Minkowski space. We will not attempt to study this here.

In the rest of the paper, we first review the null orbifold and its singular metric in section 2. In section 3, after describing the latter in the Chern-Simons formulation, we observe that its holonomy has trivial eigenvalues, and resolve the parabolic pinch singularity using flat space higher spin theories. In section 4 we review the 4-point tachyon amplitude of \cite{LMS} and present an exhaustive scan of its divergences in section 5. We conclude with a critical discussion of the meaning, potential and caveats involved in higher spin singularity resolution and its connection to string theory. We include three appendices: one contains an explicit matrix representation of the $\fsl(3,\RR)$ algebra generators used in the higher spin resolution, another one demonstrates that the resolved metric is a Kundt metric with vanishing polynomial curvature invariants to all orders and all degrees, and a third one deals with some technical details of the vertex operators and OPEs used in the string amplitude discussion.

\section{The Geometry of the Null Orbifold}

Consider $\bR^{2,1}$ in light cone coordinates $ X \equiv (x^\pm, x)= \left((x^0\pm x^1)/\sqrt{2}, x\right)$: 
\begin{equation}
ds^{2} = -2dx^{+}dx^{-} + dx^{2}.
\end{equation}
The null orbifold is the manifold $\bR^{2,1}/\Gamma$ obtained after the discrete identification of points in $\bR^{2,1}$ under the action $\Gamma$ generated by the Killing vector field $\zeta =e^{i\ell J}$ where $J$ is the Lie algebra generator of a null rotation
\begin{equation}
J = \frac{1}{\sqrt{2}} (J^{0x}+J^{1x})\in \ISO(2,1)
\end{equation}
It was first discussed in \cite{Horowitz} and shown to be supersymmetric in \cite{JoseJoan}. Here we follow the presentation in  \cite{Joan, LMS}.

The finite action in light-cone coordinates gives rise to the identification
\begin{equation}
\mathbf{X} = \left(
\begin{array}{ccc}
x^{+} \\
x  \\
x^{-} 
\end{array} \right) \sim \zeta  \mathbf{X} = \left(
\begin{array}{ccc}
x^{+} \\
x +\ell x^{+} \\
x^{-} + \ell x+ \frac{1}{2} \ell^{2} x^{+}  \label{nullorb}
\end{array} \right)
\end{equation}
Notice that the apparent arbitrariness in the parameter $\ell$ can always be absorbed by a boost $x^\pm\to\gamma^{\pm 1}x^\pm$. We use this fact to set $\ell=2\pi$ from now on. 

A local coordinate system that is convenient to discuss the geometry of the null orbifold is
\begin{equation}
y^{+} = x^{+}, \ \  
y = \frac{x}{x^{+}}, \ \ 
y^{-} = x^{-} - \frac{1}{2}\frac{ x^2}{x^{+}}\,.
\end{equation}
This is an adapted coordinate system, in the sense that the identification \eqref{nullorb} reduces to a shift in the new coordinates
\begin{equation}
 \left(
\begin{array}{ccc}
y^{+} \\ y  \\ y^{-} 
\end{array} \right)
\sim
\left(
\begin{array}{ccc}
y^{+} \\ y + 2\pi  \\ y^{-} 
\end{array} \right).
\end{equation}
The metric becomes, 
\begin{equation}
  ds^2 = -2 dy^{+} dy^{-} + (y^{+})^2 (dy)^2 
 \label{pinch}
\end{equation}
If we interpret $y^+$ as light cone time, the metric \eqref{pinch} describes two cones whose size depends on time. Thus, we have a contracting universe for $y^+<0$, up to $y^+=0$, where the local coordinate system breaks down corresponding to the fixed points of the null orbifold and then followed by an expanding universe for $y^+>0$.

This local description of the null orbifold \eqref{pinch} is sometimes called the parabolic pinch  \cite{LMS}. It is a geodesically incomplete spacetime whose maximal extension is the global orbifold \eqref{nullorb}. The latter is non-Hausdorff at $x^+=0$ \cite{LMS} and it has singularities at its fixed points $x^+=x=0$. In the following we will resolve the singularity at $y^+=0$ and study the divergences of the string theory $2\to 2$ tachyon scattering amplitude in the null orbifold \eqref{nullorb} in the large $\alpha^\prime$ limit.

\section{Higher Spins on the Parabolic Pinch} 
\label{sec:resolution}

Consider the parabolic pinch metric in the form
\begin{equation}
ds^2=-dT^2+dX^2+\frac{(T+X)^2}{2}dY^{2},
\label{pinchmink}
\end{equation}
where $y^\pm = (T\pm X)/\sqrt{2}$ and $y=Y$ brings it back to the form \eqref{pinch}. Since this is a quotient of $\RR^{1,2}$, it is a classical solution of d=1+2 flat space pure gravity. This theory is a Chern-Simons theory with $\ISO(2,1)$ gauge group \cite{Witten, achucarro}. It is possible to construct  a higher spin theory version of it by increasing the gauge group \cite{Arjun} so that \eqref{pinchmink} remains a classical solution.

Our strategy is to look for a classical solution to the flat connection equations of motion governing the higher spin theory while preserving the holonomy of the solution \eqref{pinchmink}. Thus, we use the principle based on the gauge invariance of the holonomy introduced in \cite{Gutperle} and further used in \cite{Maloney} for the resolution of higher spin black hole singularities. Here, we will be interested in constructing a solution in a higher spin gauge where the metric is resolved. Preservation of the holonomy guarantees we are discussing the same physical solution as the starting \eqref{pinchmink}. 

\subsection{Chern-Simons Gauge Field} 

The first step is to rewrite \eqref{pinchmink} in the language of gauge fields. Instead of working directly with the $\ISO(1,2)$ gauge theory, we will use the language of the $\SL(2)\times \SL(2)$ Grassmann valued connection introduced in \cite{Shubho2, Avinash}:
\begin{equation}
A^{\pm}=\left(\omega^{i}\pm\epsilon\: e^{i}\right)T_{i}
\end{equation}
where $\epsilon$ is a formal Grassmann parameter, $\epsilon^2=0$. $T_i$ $(i=\pm 1,0)$ are the generators of $\fsl(2,R)$ introduced in appendix \ref{sec:matrix}, $e^i$ is an orthonormal frame for the metric \eqref{pinchmink} and $\omega^i$ is the Hodge dual of the spin connection\footnote{We use the convention $\epsilon^{012}=1$.}
\begin{equation}
  \omega^i = \frac{1}{2}\epsilon^i_{jk}\omega^{jk} \quad \text{with} \quad de^i + \omega^i\,_j\wedge e^j = 0\,.
\end{equation}

The most natural orthonormal frame for the parabolic pinch metric \eqref{pinchmink} is
\begin{equation}
e^{T}  =  dT, \ \  e^{X}  =  dX, \ \  e^{Y}  =  \frac{T+X}{\sqrt{2}}dY\,.
\label{triads}
\end{equation}
This determines the spin connection $\omega^T\,_X=0,\,\omega^T\,_Y=-\omega^X\,_Y = dY/\sqrt{2}$. Equivalently,
\begin{equation}
\omega^{T}  =  -\frac{1}{\sqrt{2}}dY,\ \  \omega^{X}  =  \frac{1}{\sqrt{2}}dY,\ \ \omega^{Y}  =  0.
\label{spincon}
\end{equation}
This gives rise to the gauge field
\begin{equation}
A^{\pm}=\left(-\frac{dY}{\sqrt{2}}\pm\epsilon\: dT\right)T_{0}+\left(\frac{dY}{\sqrt{2}} \pm\epsilon\: dX\right)T_{1}\pm\epsilon\left(\:\frac{T+X}{\sqrt{2}}d Y\right)T_{2}. 
 \label{LMSconnection}
\end{equation}
 
The metric \eqref{pinchmink} has a non-trivial cycle, the Y-cycle. The holonomy of the gauge field \eqref{LMSconnection} around the Y-cycle equals,
\begin{equation}
\begin{aligned}
W_{Y}^{\pm} &\equiv  \text{P}\exp\left(\oint\: dY\: A_{Y}^{\pm}\right) \\
  &=  \exp\left[2\pi\left(-\frac{T_{0}}{\sqrt{2}}+\frac{T_{1}}{\sqrt{2}}\pm\epsilon\: \frac{T+X}{\sqrt{2}}\: T_{2}\right)\right] \equiv \exp\left[w_Y^{\pm}\right]
\end{aligned}  
\end{equation} 
We can characterise this holonomy either through its eigenvalue spectrum or using its characteristic polynomial coefficients. In the first approach, the $w_Y^{\pm}$ eigenvalues equal $(0,0,0)$. In the second approach, we use that any $\fsl(3,\RR)$ matrix (even one with Grassmann entries), such as $w_Y^{\pm}$, satisfies \cite{Maloney}
\begin{equation}
  \left(w_Y^{\pm}\right)^3 = {\rm Det} (w_Y^\pm)\, \mathbb{I} + \frac{1}{2}\text{tr}\left(w_Y^\pm\right)^2\,\left(w_Y^{\pm}\right)^2\,.
 \label{eq:chapol}
\end{equation}
The parabolic pinch is such that both invariants vanish
\begin{equation}
  {\rm Det} (w_Y^\pm) = \text{tr}\left(w_Y^\pm\right)^2 = 0\,.
\label{eq:hol}
\end{equation}
We have used the Grassmann approach of \cite{Avinash} to compute these holonomies, but a skeptic who is suspicious of the Grassmann approach and our use of the $AdS_3$ generators might want to repeat the same result using matrix representations of $\ISO(2,1)$ directly. A convenient way to do this is to use the adjoint representation matrices of $\ISO(2,1)$ that can be directly read off from the algebra. We have checked that the result is again that the eigenvalues of the holonomies are zero\footnote{Except that now there are six (instead of 3) eigenvalues. This is because effectively the $A^+$ and the $A^-$ eigenvalues show up together in this approach.}. 

This result may be surprising. A trivial C-S holonomy has sometimes been considered in the literature (see eg. \cite{Joris}) as the definition of a regular geometry. There is no contradiction though because the geometries considered in \cite{Joris} were required to have the topology of global AdS and therefore, not to have non-trivial cycles\footnote{We thank Joris Raeymeakers for a discussion on this.}. In metrics with no non-contractible cycles, this is a valid demand, but in a singular geometry with non-contractible cycles, such as ours, it is not clear whether there is a definite statement about holonomy that one can make. In the case of the Milne orbifold considered in \cite{Shubho2}, for example, it was found that the holonomy is non-trivial. Our result shows that having a trivial Chern-Simons holonomy is not a sufficient condition for regularity\footnote{Neither it is a necessary condition: one can obviously have geometries with non-trivial cycles with non-trivial holonomy around them, which are regular, see for example the resolved Milne metric in \cite{Shubho2} involving higher spin fields. It is perhaps worth investigating whether one can have regular geometries with non-trivial cycles and non-trivial holonomies in the pure spin-2 theory. Note also that having a non-trivial cycle is a necessary but not sufficient condition for non-trivial holonomy. The flat cylinder $ds^2= -dT^2 + dX^2 + dY^2$, with $Y \sim Y +2 \pi$, is a flat connection in the Chern-Simons language, but the holonomy eigenvalues are all zero.}.

\subsection{Higher spin resolution}

Our goal is to turn on spin-3 fields in the gauge connection:
\begin{equation}
A^{\prime\pm}=A^\pm+\sum_{a=-2}^{2}\left(C^{a}+\epsilon\: D^{a}\right)W_{a},
\end{equation}
where $W_a$ are the extra generators of $\SL(3,\RR)$ not in $\SL(2,\RR)$ (see appendix \ref{sec:matrix} for an explicit matrix representation) and $C^a, D^a$ is a set of 1-forms. This connection must be flat
\begin{equation}
  dA^{\prime\pm} + A^{\prime\pm}\wedge A^{\prime\pm} = 0\,,
\label{eq:flat}
\end{equation}
so that it satisfies the higher spin theory equations of motion. Furthermore, its holonomy around the Y-cycle must equal the one of the parabolic pinch.

Any solution to this problem will give rise to a metric and spin-3 fields \cite{Campoleoni}
\begin{equation}
\begin{split}
  g_{\mu\nu} = \frac{1}{2}\text{tr}\left(E_\mu E_\nu\right)\,, & \quad \quad \quad \Phi_{\mu\nu\rho} = \frac{1}{9}\text{tr}\left(E_{(\mu}E_\nu E_{\rho)}\right) \\
  E &= \frac{1}{2}\int \left(A^{\prime +}-A^{\prime -}\right)\,d\epsilon\,.
\end{split}
\end{equation}

We will not solve this problem in general. We are primarily interested in resolving the singularity at $T+X=0$ in the parabolic pinch \eqref{pinchmink}. To achieve this, we will look for solutions whose metric component $g_{YY}$ changes, while keeping the remaining components unmodified. This extra condition requires $D^a_T=D^a_X=0$ and the most general corrected metric component would be of the form
\begin{equation}
  g_{YY} = \frac{1}{2}\left(T+X\right)^2 + \frac{4}{3}(D^0_Y)^2 - 4D^1_YD^{-1}_Y +16D^2_YD^{-2}_Y\,.
\end{equation}
Assuming that $C^a$ and $D^a_Y$ are constant 1-forms, the flat connection condition \eqref{eq:flat} forces $C^a=0$. We are left to examine the holonomy condition. The new holonomy equals
\begin{equation}
  W^{\prime\pm}_Y = \exp\left[2\pi\left(-\frac{T_{0}}{\sqrt{2}}+\frac{T_{1}}{\sqrt{2}}\pm\epsilon\: \frac{T+X}{\sqrt{2}}\: T_{2}+ \epsilon\sum_{a=-2}^2 D^a_Y W_a\right)\right] \equiv \exp\left[\omega^{\prime\pm}_Y\right]
\end{equation} 
Requiring the preservation of the holonomy properties \eqref{eq:hol}
\begin{equation}
\mbox{det}\left(\omega^{\prime\pm}_{Y}\right)=0, \ \ \mbox{tr}\left(\omega^{\prime\pm}_{Y}\,^{2}\right)=0 \quad \Rightarrow \quad D^2_Y=0\,.
\end{equation}



As a particular resolution of the null orbifold, we choose $D^0_Y=9 p/2$ setting the remaining constants to zero $(D^{\pm 1}=D^{\pm 2}=0)$ for simplicity. The final configuration is
\begin{equation}
\begin{split}
  ds^2 &=-dT^2+dX^2+\left(\frac{(T+X)^2}{2}+27 p^2\right)dY^{2} \label{resmetr} \\
  \Phi_{YYY} &=  -18 p^3+p(T+X)^2, \quad \Phi_{XXY}  = -p/3, \quad \Phi_{TTY} =  p/3\,.
\end{split}
\end{equation}
Thus, in this frame, one can interpret the spin-3 fields as the matter supporting a resolved null orbifold for any $p\neq 0$. Because of this, both the Ricci and the Riemann tensors do not vanish when $p\neq 0$. Perhaps more surprisingly, one can check that $R$ and $R_{abcd}R^{abcd}$ do vanish for the resolution. In fact, an even stronger statement holds : the resolved metric \eqref{resmetr} has no non-vanishing polynomial scalar invariants constructed out of the Riemann tensor and its covariant derivatives. This is because the metric \eqref{resmetr} is a so-called Kundt metric, which is an example of a Vanishing Scalar Invariant (VSI) spacetime \cite{Coley}. We sketch the proof in Appendix~\ref{sec:kundt}. This shows that in some sense the null orbifold's resolution has a milder curvature than the one found for Milne, where these scalars were computed to be finite and non-zero everywhere. 


\section{The Four Point String Amplitude}

Having identified a possible resolution of the null orbifold in a higher spin gauge theory, we now turn our attention to the 2-to-2 string scattering amplitudes for tachyon vertex operators on the null orbifold studied in \cite{LMS}. Our goal is to provide a more exhaustive investigation of  {\em all} divergences in these amplitudes than the one existing in the literature and to study its behaviour in the large $\alpha^\prime$ limit.

The momentum space Virasoro-Shapiro amplitude is one of our main objects of study. This was computed in \cite{LMS} and equals :
\begin{equation}\label{amp1}
A_4  =  \frac{8(2\pi)^3 i g_{s}^2}{\alpha^{'}} \int \left(  \prod_{i=1}^4 \frac{dp_i}{\sqrt{2\pi p_{i}^{+}}} \right) \delta(p_1 + p_2 - p_3 - p_4)  e^{ i F} \delta (E) A(s,t).
\end{equation}
This expression suppresses a factor of $(2\pi)^{24} \delta(p^{+}_1 + p^{+}_2 - p^{+}_3 - p^{+}_4) \delta(\vec{p}_{\perp 1} + \vec{p}_{\perp 2} - \vec{p}_{\perp 3} - \vec{p}_{\perp 4})$ dealing with part of the momentum conservation. Furthermore,
\begin{equation}
  E = p^{-}_1 + p^{-}_2 - p^{-}_3 - p^{-}_4 \quad \text{and} \quad  F = p_1 \xi_1+ p_2 \xi_2- p_3 \xi_3 - p_4 \xi_4,
\end{equation}
with $\xi_i \equiv - J_i / p^{+}_i$ and $J_i$ stands for the $\U(1)$ orbifold invariant charge associated to the operator $\hat{J}=-i\left(x^+\partial_x + x\partial_{x^-}\right)$. $A(s,t)$ is the same quantity defined by $A(L_s,L_t,L_u)$ in \eqref{eq:multgam}.

It was shown in \cite{LMS} that the momentum space Virasoro-Shapiro amplitude \eqref{amp1} can be reduced to a single integral, 
\begin{equation}
A_4  =  \frac{8(2\pi)^2 i g_{s}^2}{\alpha^{'}} \delta(J_1 + J_2 - J_3 - J_4) \int_{- \infty}^{\infty} \frac{dq}{|q|} \exp\left[ \frac{i}{2} \left( q \xi_{-} + \frac{\alpha \xi_{+} }{q} \right)   \right]  A(L_s,L_t,L_u)
 \label{ampsimp}
\end{equation}
where
\begin{equation}
A(L_s,L_t,L_u) = \pi \frac{\Gamma(-\frac{\alpha^{'}}4{}L_s)\Gamma(-\frac{\alpha^{'}}4{}L_t)\Gamma(-\frac{\alpha^{'}}4{}L_u)}{\Gamma(1+\frac{\alpha^{'}}{4} L_s)\Gamma(1+\frac{\alpha^{'}}{4} L_t)\Gamma(1+\frac{\alpha^{'}}{4} L_u)}
\label{eq:multgam}
\end{equation}
with $L_s$, $L_t$ and $L_u$ being the standard Mandelstam invariants
\begin{equation}
\begin{aligned}
L_s &= s - m^2 + i\epsilon = (p^{+}_1 + p^{+}_2)  \left( q_{+}^2 + \frac{m_{1}^{2}}{p_{1}^{+}} + \frac{m_{2}^{2}}{p_{2}^{+}} \right) - m^{2}_s + i \epsilon, \\
L_t &= t- m^2 + i\epsilon \\
 & = (p^{+}_3 - p^{+}_1)  \left(\frac{m_{3}^{2}}{p_{3}^{+}} - \frac{m_{1}^{2}}{p_{1}^{+}} \right) - m^{2}_t  -\mu_{12}  \left(  \sqrt{\frac{p^{+}_3}{p^{+}_1}} q_{+} - \sqrt{\frac{p^{+}_4}{p^{+}_2}} q_{-}  \right)^2   +  i \epsilon, \\ 
L_u &= u-m^2 + i\epsilon \\
&= (p^{+}_4 - p^{+}_1)  \left(\frac{m_{4}^{2}}{p_{4}^{+}} - \frac{m_{1}^{2}}{p_{1}^{+}} \right) - m^{2}_u  -\mu_{12}  \left(  \sqrt{\frac{p^{+}_4}{p^{+}_1}} q_{+} + \sqrt{\frac{p^{+}_3}{p^{+}_2}} q_{-}  \right)^2   +  i \epsilon\,.
\end{aligned}
\end{equation}
The string amplitude \eqref{ampsimp} and Mandelstam invariants have already been written in terms of the variables and parameters :
\begin{equation}
\begin{aligned}
\xi_{\pm} &= \sqrt{\mu_{12}} ( \xi_1 -\xi_2 ) \pm \sqrt{\mu_{34}} (\xi_3 - \xi_4 ), \quad q_{\pm} = \frac{1}{2} \left( q \pm \frac{\alpha}{q} \right), \\ 
\mu_{ij} &= \frac{p^{+}_{i}  p^{+}_{j}}{p^{+}_{i}+ p^{+}_{j}}, \quad \left(i\neq j\right) \quad \alpha =\frac{m_{3}^{2}}{p_{3}^{+}} + \frac{m_{4}^{2}}{p_{4}^{+}}- \frac{m_{1}^{2}}{p_{1}^{+}} - \frac{m_{2}^{2}}{p_{2}^{+}}, \\ 
m^{2}_{i} &= m^2 + \vec{p}^{\,2}_{\perp i} \quad i=1,2,3,4 \\
m_{s}^2 &=  m^2 + (\vec{p}_{\perp 1} + \vec{p}_{\perp 2})^2,\quad m_{t}^2 =  m^2 + (\vec{p}_{\perp 1} - \vec{p}_{\perp 3})^2, \quad
m_{u}^2 =  m^2 + (\vec{p}_{\perp 2} - \vec{p}_{\perp 3})^2
\end{aligned}
\end{equation}

We refer the reader to \cite{LMS} for a proper discussion on the physical meaning of the different variables defined above. See also our Appendix \ref{sec:vertex}. For our purposes, it is sufficient to identify how many of them are independent, so that we can perform a complete scan for amplitude divergences in the entire parameter space. The independent kinematic parameters are : $p^+_i,\,J_i,\,p_i,\,\vec{p}_{\perp i}$ $i=1,2,3$. This is because the fourth particle data is always given by momentum conservation and our external particles are on-shell tachyons fixing $p^-_i$ and $m^2$ to be
\begin{equation}
p_{i}^{-}= \frac{p_{i}^2 + m_i^2}{2 p_{i}^{+}}, \ \ {\rm where} \ \ m_i^2=m^2+(\vec{p}_{\perp i})^2, \ \ {\rm with} \ \ m^2 = - \frac{4}{\alpha^\prime}.
\end{equation}
The physical amplitude \eqref{ampsimp} only depends on 12 of these : $p^+_i,\,J_i$ $i=1,2,3$, the three transverse momenta magnitudes $(\vec{p}_{\perp i})^2$ and their relative scalar products that we shall denote as
\begin{equation}
P_{12} = \vec{p}_{1\perp}.\vec{p}_{2\perp}\,, \,\,P_{13} = \vec{p}_{1\perp}.\vec{p}_{3\perp}\,, \ \,P_{23} = \vec{p}_{2\perp}.\vec{p}_{3\perp}.
\end{equation}
Notice the last six parameters fix $m_i^2$ together with $m_s$, $m_t$ and $m_u$. These must satisfy the following constraints arising from triangle inequalities and momentum conservation
\begin{equation*}
\begin{aligned}
\left(\vec{p}_{1\perp}\right)^2+\left(\vec{p}_{2\perp}\right)^2 \ge 2|P_{12}| &\Longleftrightarrow  |P_{12}| \le \frac{m_1^2 + m_2^2}{2} + 4, \nonumber\\
\left(\vec{p}_{1\perp}\right)^2+\left(\vec{p}_{3\perp}\right)^2 \ge 2|P_{13}| &\Longleftrightarrow  |P_{13}| \le \frac{m_1^2 + m_3^2}{2} + 4 ,\nonumber\\
\left(\vec{p}_{2\perp}\right)^2+\left(\vec{p}_{3\perp}\right)^2 \ge 2|P_{14}| &\Longleftrightarrow  |P_{23}| \le \frac{m_2^2 + m_3^2}{2} + 4, \nonumber\\
\vec{p}_{1\perp}+\vec{p}_{2\perp} = \vec{p}_{3\perp}+\vec{p}_{4\perp} &\Longleftrightarrow  m_4=\sqrt{m_1^2+m_2^2+m_3^2+8+2 P_{12}-2 P_{13}-2 P_{23}}. \nonumber
\end{aligned}
\end{equation*}
 Furthermore, the Mandelstam invariants satisfy the identify
\begin{equation}
  \left(L_s + L_t + L_u\right)\alpha^\prime = -4\,,
 \label{eq:manid1}
\end{equation}
and all the momenta $p^+_i > 0$ since all on-shell particles carrying positive energy have this property in the light cone gauge. In all our explicit numerical evaluations of integrals and plots presented below, we have checked all these conditions are always satisfied.

Note that this is the count of dimensionless parameters. This is because $\alpha'$ is also dimensionful and can be used to make these parameters dimensionless. Conversely, since the background is an orbifold of flat space and has no dimensionful parameters, a useful way to characterize the large $\alpha^\prime$ limit will be via large momentum transfer.

\section{Divergences of the Four-point function}
\noindent
To scan all the divergences in the four-point amplitude \eqref{ampsimp}, we must study
\begin{itemize}
\item if the integr{\bf and} diverges somewhere on the integration domain too badly,
\item if the integr{\bf al} fails to converge at one of its boundaries.
\end{itemize}
We will set  $\alpha^\prime=1$ for convenience and $\xi_{i}=0=J_i$ since the latter only affects the phases in the integrand and will not modify its divergent structure. In this case, the amplitude  \eqref{ampsimp} is invariant under $q\to -q$. Thus, we will only analyse the positive $q$ half-line. 

The multiple gamma functions in \eqref{ampsimp} have poles in the integrand. As explained in \cite{BenAyush}, integration around these poles is finite because their Cauchy Principal Value (CPV) is finite, except when the poles lie at the boundary of the integration region or they are extremal values of the continuous argument of the gamma function. Thus, there are three different situations in which divergences appear :
\begin{enumerate}
\item the integrand blows-up in the $q\to 0$ or $q\to \infty$ boundary limits \& $\nexists$ poles in the gamma functions.
\item gamma function poles at the boundary of the integral. 
\item gamma function poles when their continuous argument takes an extremal value.
\end{enumerate}

\subsection{Boundary divergences unrelated to gamma poles}
\label{eq:genkin}

Consider the four-point string amplitude \eqref{ampsimp} for generic kinematic configurations. Since in the light-cone gauge all momenta $p_i^+ > 0$, all Mandelstam invariants $L_s$, $L_t$, and $L_u$ diverge (see Fig. 1) when $q\to \infty$. Analytically, these invariants scale like
\begin{equation}
L_s \to 4 C_s\ q^2 \,, \quad L_t  \to -4\, C_{t}\ q^2\,, \quad  L_{u} \to -4\, C_{u}\ q^2 \quad \quad q\to \infty
\end{equation}
where
\begin{equation*}
C_s =\frac{(p_{1}^{+}+p_{2}^{+})}{16} \,, \,\,\, C_t =  \frac{\left(\sqrt{p_2^+ p_3^+}-\sqrt{p_1^+ p_4^+}\right){}^2 }{16 \left(p_1^++p_2^+\right)}\,,\,\,\,
 C_{u} =  \frac{\left(\sqrt{p_1^+ p_3^+}+\sqrt{p_2^+ p_4^+}\right){}^2}{16 \left(p_1^++p_2^+\right)}\,.
\end{equation*}
\begin{figure}[h]
\centering
\includegraphics[width=0.8\textwidth]{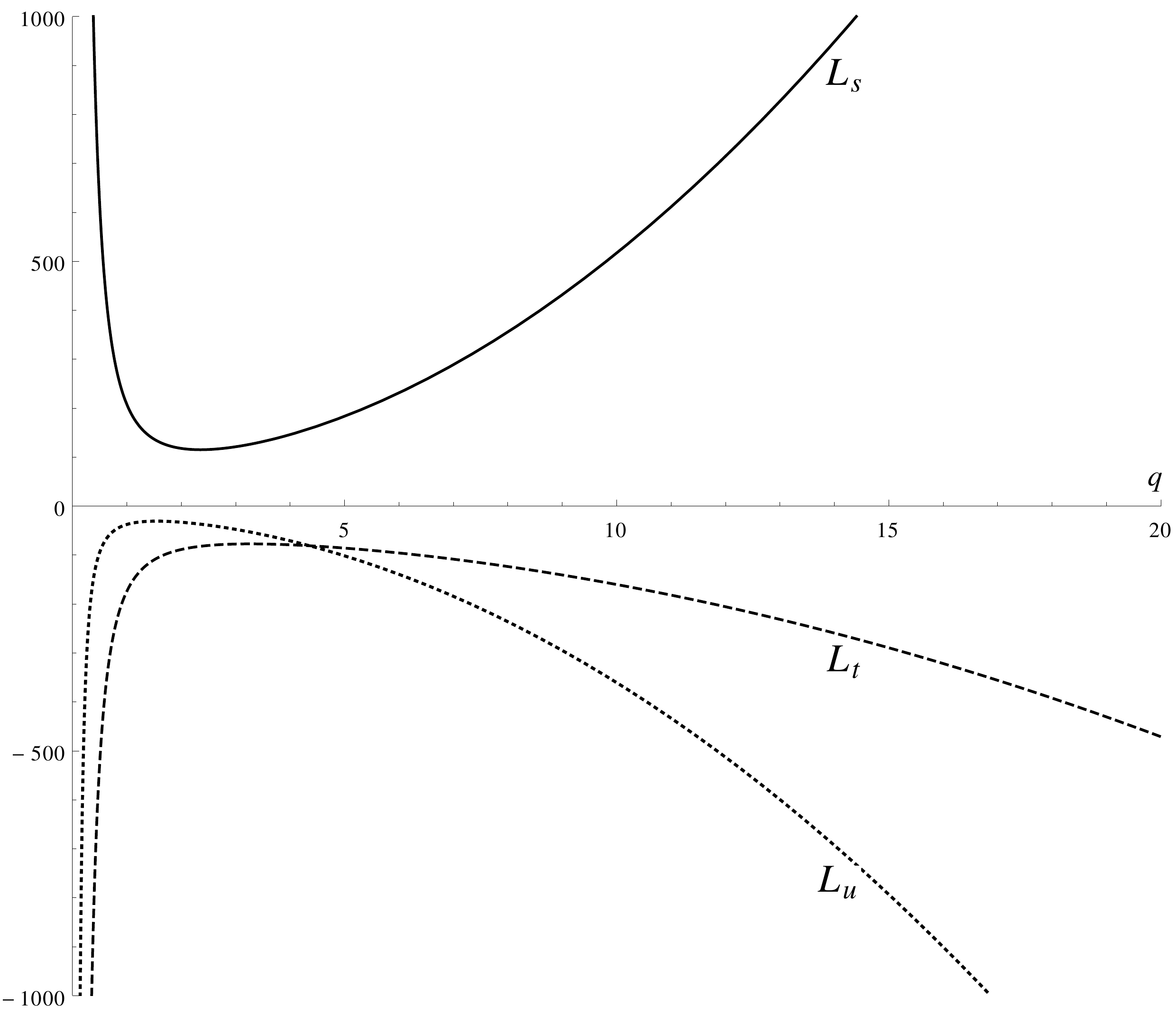}
\caption{Plot of $L_s,L_t$ and $L_u$ for generic case in positive $q$. Values of kinematic parameters used for the plot are $m_1=4.0, m_2=3.0, m_3=2.5, p_1^+=2.0, p_2^+=16.0, p_3^+=10.0, P_{12}=5.0, P_{13}=4.0, P_{23}=11.0$}
\label{fig:circles}
\end{figure}
Notice that $\left(L_s + L_t + L_u\right)\alpha^\prime = -4$ reduces to $C_s-C_t-C_u=0$ in this limit, which is equivalent to the momentum conservation $p_{1}^{+} +p_{2}^{+} = p_{3}^{+} + p_{4}^{+}$. Using these simplifications and the Stirling approximation to evaluate the gamma functions in the large $q$ limit, the four-point amplitude becomes
\begin{equation}
  A_4 \propto \int_{}^{\infty} dq \, q^{-7}\, B^{-q^2} \quad \text{where} \quad B=C_{s}^{2 C_{s}} \, C_{t}^{-2 C_{t}} \, C_{u}^{-2 C_{u}}\,.
\end{equation}
Using momentum conservation, i.e. $C_s-C_t-C_u=0$, it can be shown that $B>1$. Thus, for generic kinematic configurations the string amplitude \eqref{ampsimp} does not diverge in the $q \to \infty$ limit. 

Similarly, it is easy to argue that the same generic convergent behaviour exists in the $q \to 0$ limit. One way to see this is to realise that 
$q_\pm \to \alpha^2/(4q^2)$ when $q\to 0$. Thus, the leading contribution to the integrand independent from the measure factor $dq/|q|$ can be computed by the map $q^2\to \alpha^2/q^2$. This argument fixes the $q$ scaling of the integrand to be 
\begin{equation}  
  A_4 \propto \int_0 \frac{dq}{|q|} q^6\,B^{-\alpha^2/q^2}\,. 
\end{equation} 
There is indeed no divergence in this case either.

 
Our discussion above dealt with generic kinematic configurations where all Mandelstam invariant diverge in the boundaries of the q-space. In the following subsections, we discuss particular kinematic conditions where some of the Mandelstam invariants approach finite values in these boundary limits $\left(q\to 0,\,\infty\right)$.

\subsubsection{Case $p_{1}^{+}=p_{3}^{+}$ in the $q \to \infty$ limit}
\label{sec:p13}
\noindent
For this particular kinematic configuration, $L_t$ remains finite and the $q\to \infty$ limit equals (see Fig. 2)
\begin{equation}
\begin{aligned}
L_s & \to \frac{(p_{1}^{+}+p_{2}^{+})}{4} q^2 \to \infty \,\,\, , \,\, L_t \to -m_{1}^{2}-m_{3}^{2}-4 +2 P_{13} \\
 L_{u} & \to -\frac{(p_{1}^{+}+p_{2}^{+})}{4} q^2 \to -\infty
\end{aligned}
\end{equation}
Thus, this corresponds to the Regge limit condition $L_{s} \to \infty$ with $L_t$ fixed. In this regime, $A(L_s,L_t,L_u)$ reduces to 
\beq
 A(L_s,L_t,L_u) \to \pi \Big(\frac{L_s}{4}\Big)^{\frac{L_t}{2}} \, \frac{\Gamma(-\frac{L_t}{4})}{\Gamma(1+\frac{L_t}{4})}\nonumber 
\eeq
Using these results, the four-point function \eqref{ampsimp} behaves, close to the boundary $q\to\infty$, as
\begin{equation*}
  A_4 \propto \int_{}^{\infty} dq \bigg(\frac{1}{q}\bigg)^{(1-L_t)}\,.
\end{equation*}
\begin{figure}[h]
\centering
\includegraphics[width=0.8\textwidth]{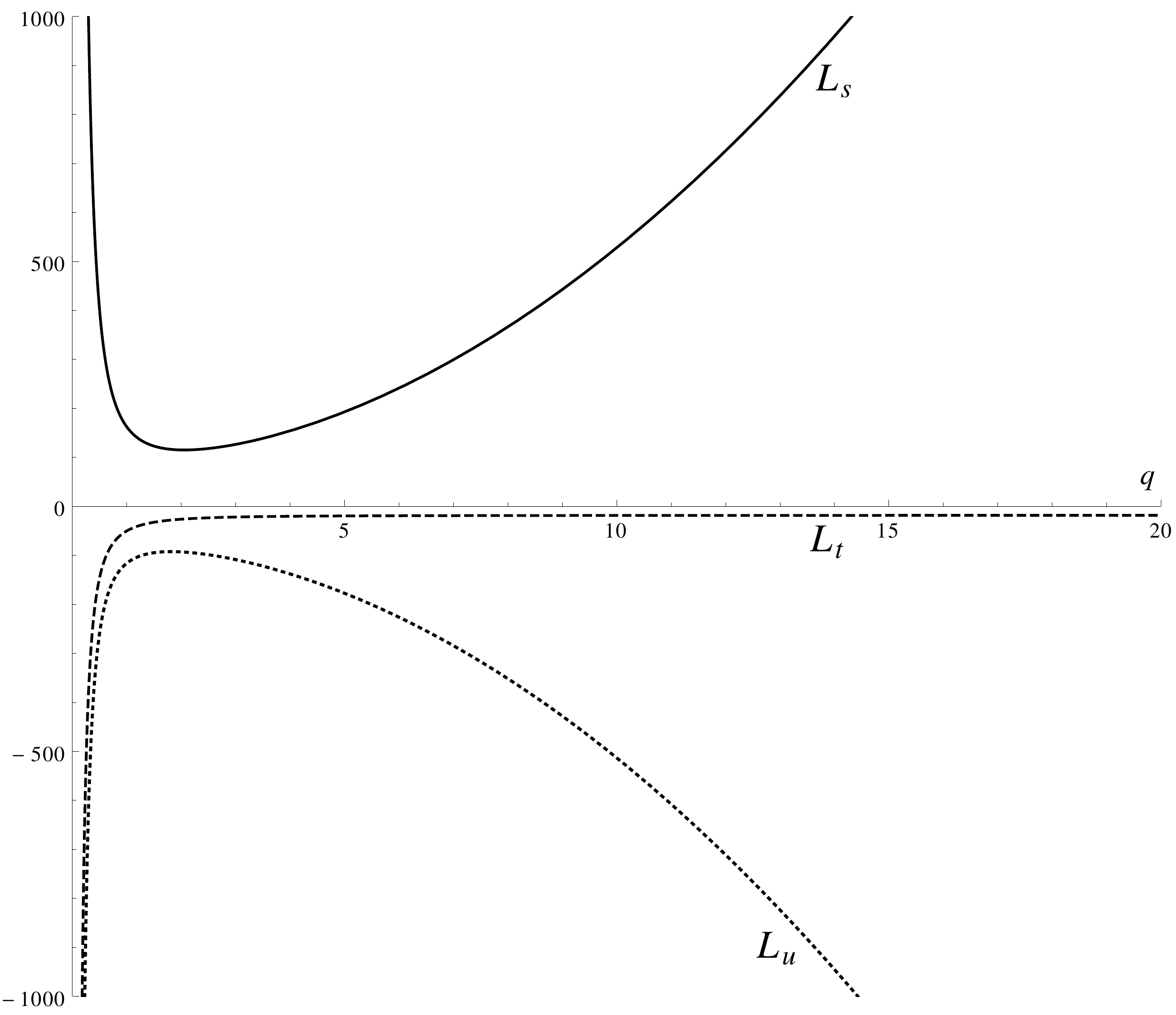}
\caption{Plot of $L_s,L_t$ and $L_u$ for $p_{1}^{+}=p_{3}^{+}$ in positive $q$. Values of the kinematic parameters used for the plot are $m_1=4.0, m_2=3.0, m_3=2.5, p_1^+=2.0, p_2^+=16.0, p_3^+=2.0, P_{12}=5.0, P_{13}=4.0, P_{23}=11.0$.}
\label{fig:circles}
\end{figure}
This integral diverges whenever 
\beq
(1-L_t) \le 1 \,\, \Longleftrightarrow  \,\, m_{1}^{2}+m_{3}^{2}+4 -2 P_{13} \le 0 \,\, \Longleftrightarrow  \,\, (\vec{p}_{1\perp}-\vec{p}_{3\perp})^2\alpha^\prime \le 4 
\eeq
where we reinserted the dependence on $\alpha^\prime$ in the last inequality.

\subsubsection{Case $p_{2}^{+}=p_{3}^{+}$ in the $q \to 0$ limit}
\noindent
This is a similar kinematic configuration to the one above in which we keep $L_u$ finite. We find that in the $q\to 0$ limit, we have (see Fig. 3)
\begin{equation}
\begin{aligned}
L_s &\to \frac{(p_{1}^{+}+p_{2}^{+})\,\alpha^2}{4} q^{-2} \to \infty \,\,\, , \,\, L_t \to -\frac{(p_{1}^{+}+p_{2}^{+})\,\alpha^2}{4} q{^{-2}} \to - \infty \\
 L_{u} &\to  -m_{2}^{2}-m_{3}^{2}-4 +2 P_{23} 
\end{aligned}
\end{equation}
Again, we have the Regge limit condition $L_{s} \to \infty$ and $L_u$ fixed. In this regime, $A(L_s,L_t,L_u)$ reduces to
\begin{equation*}
 A(L_s,L_t,L_u) \rightarrow \pi \Big(\frac{L_s}{4}\Big)^{\frac{L_u}{2}} \, \frac{\Gamma(-\frac{L_u}{4})}{\Gamma(1+\frac{L_u}{4})}
\end{equation*}
Thus, the four-point amplitude behaves, close to the function integral is found to go as
\begin{equation*}
  A_4 \propto \int_{0}^{} dq \bigg(\frac{1}{q}\bigg)^{(1+L_u)}
\end{equation*}
\begin{figure}[h]
\centering
\includegraphics[width=0.8\textwidth]{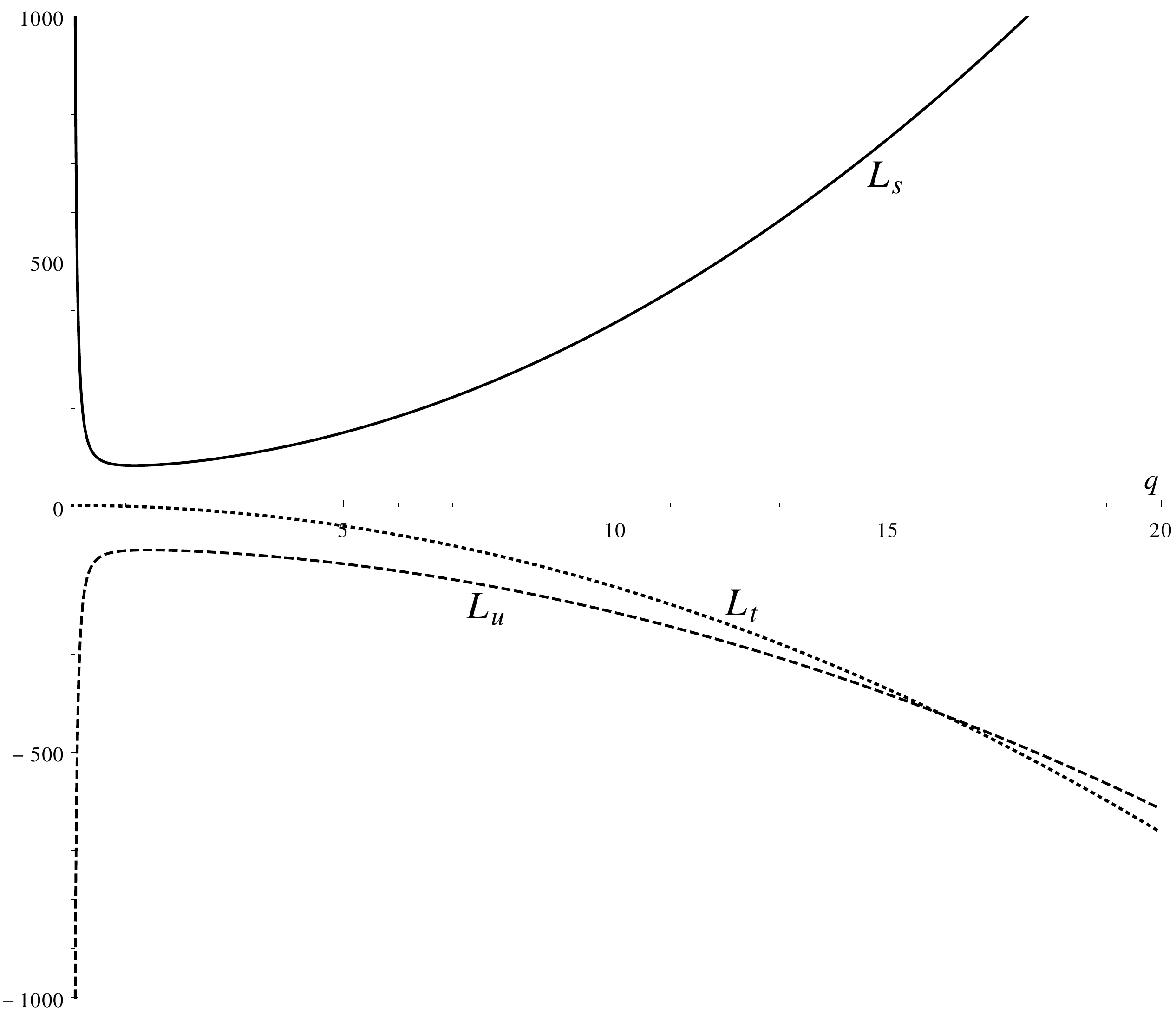}
\caption{Plot of $L_s,L_t$ and $L_u$ for $p_{2}^{+}=p_{3}^{+}$ in positive $q$. The kinematic parameters used for the plot are $m_1=4.0, m_2=3.0, m_3=2.5, p_1^+=2.0, p_2^+=10.0, p_3^+=10.0, P_{12}=5.0, P_{13}=4.0, P_{23}=11.0$.}
\label{fig:circles}
\end{figure}
This integral diverges whenever 
\begin{equation}
(1+L_u) \ge 1 \,\, \Longleftrightarrow  \,\, m_{2}^{2}+m_{3}^{2}+4 -2 P_{23} \le 0 \,\, \Longleftrightarrow  \,\, (\vec{p}_{2\perp}-\vec{p}_{3\perp})^2\alpha^\prime \le 4
\end{equation}
where we reinserted the dependence on $\alpha^\prime$ in the last inequality.

\subsubsection{Case $\alpha=0$}
\label{sec:alpha}
\noindent
When $\alpha=0$, the $q\to 0$ boundary limit keeps all the Mandelstam invariants finite (see Fig. 4)

\begin{figure}[h]
\centering
\includegraphics[width=0.8\textwidth]{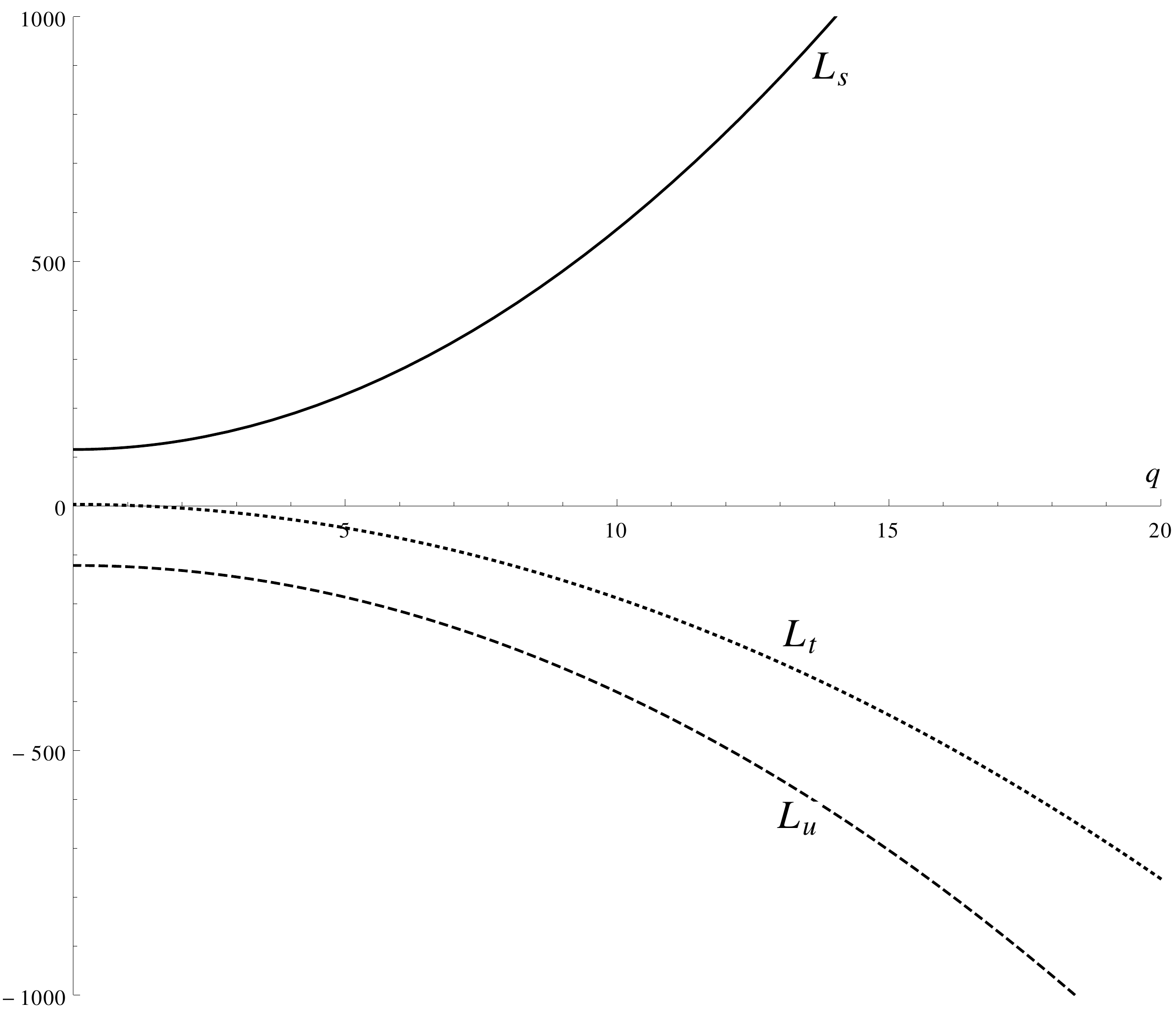}
\caption{Plot of $L_s,L_t$ and $L_u$ for $\alpha =0$ in positive $q$. The kinematic parameters used for the plot are $m_1=4.0, m_2=3.0, m_3=2.5, p_1^+=2.0, p_2^+=16.0, p_3^+\approx 15.6418.., P_{12}=5.0, P_{13}=4.0, P_{23}=11.0$. The approximate value of $p_3^+$ is a consequence of demanding $\alpha=0$.}
\label{fig:circles}
\end{figure}

\begin{equation}
\begin{aligned}
L_s &\to \frac{m_1^2 p_{2}^{+}}{p_{1}^{+}}+\frac{m_2^2 p_{1}^{+}}{p_{2}^{+}}-2 P_{12}-4 \,\,\, , \,\,
L_t \to -\frac{m_1^2 p_{3}^{+}}{p_{1}^{+}}-\frac{m_3^2 p_{1}^{+}}{p_{3}^{+}}+2 P_{13}-4 \\
 L_{u} &\to  -\frac{m_1^2 p_{4}^{+}}{p_{1}^{+}}-\frac{m_4^2 p_{1}^{+}}{p_{4}^{+}}+m_1^2-m_2^2-m_3^2+m_4^2+2 \text{P}_{23}-4  
\end{aligned}
\end{equation}
\noindent
The four-point amplitude behaves as
\begin{equation*}
  A_4 \propto \int_{0}^{} dq \bigg(\frac{1}{q}\bigg)\,.
\end{equation*}
Thus, it always diverges in this $q \to 0$ limit. 

In Appendix \ref{sec:vertex} we show that the $\alpha=0$ divergences can be understood as IR divergences. It is worth mentioning here that the divergence in this case is in many ways analogous to the IR divergences discussed in the Milne orbifold case in section (4.4) of \cite{BenAyush} (the so-called type-4 divergences there). They were also log divergences that occurred at the lower boundary of the integral for a specific kinematic configuration (closely analogous to the $\alpha=0$ condition considered here). This is an observation that we will come across repeatedly: despite the fact that the details in both orbifolds are quite different, 
there is a very close parallel between the Milne and null orbifold amplitudes. This is perhaps not surprising in hindsight, but it would nonetheless be nice to formalise these similarities in terms of strings in the covering space and (timelike) orbifold actions.

\subsection{Extrema Pole Divergences}
\label{extremapole}

To identify these divergences, we first determine the extrema of $L_s$, $L_t$, and $L_u$ in the positive $q$ half-line. Labelling each of these by
$q_s$, $q_t$, and $q_u$, respectively, they are given by
\begin{equation}
\begin{aligned}
q_s^2 &=  \pm \, \alpha \\
q_t^2 &= \pm \, \alpha\,\Bigg(\frac{\sqrt{p_{2}^{+} p_{3}^{+}} + \sqrt{p_{1}^{+}p_{4}^{+}}}{\sqrt{p_{2}^{+}p_{3}^{+}} - \sqrt{p_{1}^{+} p_{4}^{+}}} \Bigg) \\
q_u^2 &= \pm \, \alpha \Bigg(\frac{\sqrt{p_{2}^{+} p_{4}^{+}} - \sqrt{p_{1}^{+}p_{3}^{+}}}{\sqrt{p_{1}^{+}p_{3}^{+}} + \sqrt{p_{2}^{+} p_{4}^{+}}} \Bigg)\,.
\end{aligned}
\label{eq:extrema}
\end{equation}
We must choose the appropriate branch for each kinematic configuration to keep these values real valued. The $L_s$, $L_t$, and $L_u$ extremal values at the positive root equal
\begin{equation}
\begin{split}
  L_s(q_s) &= \left(p^{+}_1 + p^{+}_2\right)\left(\frac{m_{3}^{2}}{p_{3}^{+}} + \frac{m_{4}^{2}}{p_{4}^{+}} \right) - m^{2}_s 
  , \\
L_t(q_t) &= (p^{+}_3 - p^{+}_1)  \left(\frac{m_{3}^{2}}{p_{3}^{+}} - \frac{m_{1}^{2}}{p_{1}^{+}} \right) - m^{2}_t  
, \label{poleL}\\
L_u(q_u) &= (p^{+}_4 - p^{+}_1)  \left(\frac{m_{4}^{2}}{p_{4}^{+}} - \frac{m_{1}^{2}}{p_{1}^{+}} \right) - m^{2}_u 
. 
\end{split}
\end{equation}
Divergences in the 4-pt function amplitude \eqref{ampsimp} will occur whenever there are poles in the multiple gamma functions appearing in \eqref{eq:multgam}. These occur whenever
\begin{equation}
  L_s(q_s)=4 n, \quad \text{or} \quad L_t(q_t)= 4 m \quad \text{or} \quad  L_u(q_u)=4m^\prime \quad \text{for} \quad n,m,m^\prime \in \{0\}
 \cup \bZ^+
\end{equation}
The fact that they are indexed by integers suggests that they correspond to the tower of string states going on-shell. If so, they would be associated to IR divergences. These conditions become constraints between the kinematic parameters, but it is easy to see (``by inspection") that there exist solutions to these constraints for any non-negative integer. 

To show that these divergences indeed have an IR interpretation, we start with the vertex operators in the form (\ref{expvert}). Far away from the singularity ($x^+ \rightarrow \infty$), these take the form
\begin{equation}
V_{p^+_i,J_i}  \sim 
\exp \left[-ip_i^+x^--i{m_i^2\over 2p_i^+} x^+ 
 + i \vec{p}_{\perp i} \cdot \vec{x}_{\perp} \right] \label{vertemp}
\end{equation}
It is straightforward to see that the divergences at (\ref{poleL}) arise from the poles in the propagator in this limit: one reproduces expressions (\ref{poleL}) when one acts with the d'Alembertian $(-2\partial_-\partial_++\partial_x^2+\partial_\perp^2)$ on pairs of vertex operators\footnote{Appropriately complex-conjugated if necessary, in the appropriate channel.} of the form (\ref{vertemp}). See also the related discussions in Appendix \ref{sec:vertex}. 


\subsection{Boundary Pole Divergences}\label{boundarypole}

To study the amplitude divergences due to gamma function poles occurring at the boundary in $q$-space, we distinguish between $q\to \infty$ and $q\to 0$. For the generic kinematical configurations studied in subsection \ref{eq:genkin} with $q\to\infty$, the only gamma function poles $\alpha^\prime L_i = 4n_i$ that can occur are for $L_s$, since both $L_u$ and $L_t$ are negative. 
 For the particular configurations $p_{1}^{+}=p_{3}^{+}$ studied in subsection \ref{sec:p13}, we see that $L_t \rightarrow 4-\left(\vec{p}_{1\perp}-\vec{p}_{3\perp}\right)^2$ as $q\to \infty$, which for divergent poles 
  reduces to 
\begin{equation}
  \left(\vec{p}_{1\perp}-\vec{p}_{3\perp}\right)^2\alpha'=4(1-n_t)\,.
\end{equation}
Since $n_t$ is zero or a positive integer, the only consistent solutions are $n_t=0$, corresponding to $L_t=0$ or $n_t=1$, corresponding to $\vec{p}_{1\perp}=\vec{p}_{3\perp}$. One may think the latter condition is equivalent to trivial scattering but this is not the case. Indeed, even though $\vec{p}_{1\perp}=\vec{p}_{3\perp}$ and $p_{1}^{+}=p_{3}^{+}$, we do have
\begin{equation}
  p^-_1-p_3^- = \frac{(p_1-p_3)(p_1+p_3)}{2p_3^+}\,.
\end{equation}
Thus, non-trivial momentum exchange along the orbifold direction is allowed. Notice this case corresponds to $\alpha=0$. However, this divergence is to be distinguished from the generic $\alpha=0$ divergences we identified earlier. The divergence here exists only for more restrictive kinematics ($\vec{p}_{1\perp}=\vec{p}_{3\perp}$ and $p_{1}^{+}=p_{3}^{+}$), and they arise in the $q\to \infty$ limit, from a pole in the gamma function. 

The boundary $q=0$ has a similar discussion to the one above when $p_2^+=p_3^+$ but now involving $L_u$ giving us the following condition for the existence of divergent poles :
\begin{equation}
  \left(\vec{p}_{2\perp}-\vec{p}_{3\perp}\right)^2\alpha^\prime=4(1-n_u)\,.
\end{equation} 
The same kind of solutions $n_u=0, 1$ exists, i.e. exchanging particles $1\leftrightarrow 2$. Again, the $n_u=1$ case is a divergence that arises in the $\alpha=0$ case, and again, it is to be distinguished from the generic $\alpha=0$ divergences: the $n_u=1$ divergence arises from a gamma function pole. But unlike the $n_t=1$ divergence above which is at $q=\infty$, this one causes an enhancement of the divergence at the $q=0$ boundary.

Analogous to the extrema pole divergences of the previous section, now we show that these divergences can also be given a particle-going-on-shell interpretation and are therefore an IR effect. For concreteness we consider the $q\rightarrow \infty$ case. We first note that the vertex operators \ref{expvert}, in the $x^+ \rightarrow 0$ limit tend to 
\bea
V_{p^+_i,J_i}  \sim {1 \over \sqrt{ix^+}}\ \exp \left[-ip_i^+\left(x^--  
{(x-\xi_i)^2 \over 2 x^+}\right)+ i \vec{p}_{\perp i} \cdot \vec{x}_{\perp} \right] 
\eea
The boundary pole divergence in the $q\rightarrow \infty$ case happens when $p_1^+ =p_3^+$. When this happens, the $\Box$ operator in the $u$-channel in this limit simplifies to $\left(\vec{p}_{1\perp}-\vec{p}_{3\perp}\right)^2$ as can be checked by acting with the d'Alembertian on the vertex operator pair $V_{p_1} V_{p_3}^*$.

Typically one thinks of divergences arising from poles in the propagator in a string amplitude as an IR phenomenon: it correponds to an on-shell particle that propagates for long distances in spacetime \cite{Polchinski}. However, in our discussion of the previous paragraph there is a subtlety. This is because the form of the propagator arose from looking at the vertex operators in the near-singularity region ($x^+ \to 0$). Typically one thinks of the singularity as the ``UV", so one can think of this as a UV-enhancement of an IR divergence\footnote{We thank Ben Craps for a discussion on this, and for pointing out a different UV-enhanced IR divergence in the nullbrane \cite{LMS2}.} or as a form of UV/IR mixing. Since the pole type on-shell divergences are expected on physical grounds, this will not bother us. An entirely parallel singularity structure was found also in the Milne orbifold \cite{BenAyush}.

Before leaving this section, we also note that when $\alpha=0$, the extrema pole divergences \eqref{eq:extrema} found in the previous subsection end up moving to the boundary. But we will not double-count them as boundary pole divergences.

\subsection{List of Divergences}

We list the various divergences here for convenience:

\begin{itemize}
\item The UV divergences that happen when $p_1^+=p_3^+$ and $(\vec{p_{1\perp}}-\vec{p_{3\perp}})^2 \alpha' \le 4$, and when  $p_2^+=p_3^+$ and $(\vec{p_{2\perp}}-\vec{p_{3\perp}})^2 \alpha' \le 4$. These were found in \cite{LMS}.

\item The IR divergence that arises when $\alpha=0$.

\item Divergences arising from  $p_1^+=p_3^+$ and $(\vec{p_{1\perp}}-\vec{p_{3\perp}})^2 \alpha' = 4\ {\rm or}\ 0$, and when  $p_2^+=p_3^+$ and $(\vec{p_{2\perp}}-\vec{p_{3\perp}})^2 \alpha'= 4\ {\rm or}\ 0$. These are to be interpreted as tachyons and massless string states going on-shell. These are pole divergences (and therefore are of IR-type and physical), but they get contributions from near the singularity. 

\item Divergences from the tower of string states going on-shell discussed in subsection \ref{extremapole}.
\end{itemize}

The UV divergences that were already noted in \cite{LMS} go away when the dimensionless $\alpha^\prime$ is large enough. All the new divergences are physical IR-type divergences. 

\section{A Paradigm for Singularity Resolution}

We conclude this paper with some comments and caveats about singularity resolution in higher spin theories and their connection to
string theory.

\begin{itemize}
\item The basic philosophy being followed is that the Chern-Simons gauge theory contains a master field capturing all the information about the metric and the higher spin fields. In that sense, it parallels "classical" string field theory and it also captures tree level effects.

\item Singularities which are resolvable by gauge transformation are to be thought of as regular geometries with cycles, albeit in a gauge where the metric looks singular. The higher spin gauge transformation is to be thought of as taking the system away from this gauge so that the metric is regular.

\item We emphasize that the Chern-Simons gauge field is regular everywhere. Thus, we would not expect any issues related to quantization and singular Jacobians, though the metric can be in singular gauges. On the other hand, the connection between the Chern-Simons theory and higher spin gravity is perhaps best thought of as a classical equivalence. It is known that the spin-2 gravity theory to $\SL(2)$ C-S theory relation is probably unreliable beyond the classical regime \cite{Witten2}: one reason for this is that C-S theory is effectively topological, so one does not expect to find enough states in its spectrum to explain the enormous degeneracy necessary to explain black hole entropy in the gravity theory. 

\item The biggest caveat in this construction is that the higher spin gauge transformation, while it removes the zero of the metric, introduces zeroes in the higher spin field. Since there is no higher spin generalization of Riemannian geometry, it is not immediately clear whether this is problematic or not. We chose to be optimistic, because of a few reasons- (a) The gauge field, viewed as a master field, is regular everywhere and has a well-defined holonomy (and as we saw earlier, it can even be trivial). (b) The string amplitudes are better behaved at larger $\alpha^\prime$ as we have seen. (c) That singularities in the metric are a gauge artifact, in a theory where the metric is a gauge {\em variant} quantity seems to us as a logical possibility that is worthy of exploration. In particular, the idea that a non-trivial cycle can look like a pinch-off geometry in a gauge where the metric is singular seems plausible to us. String theory contains an enormous gauge invariance on the worldsheet in the form of conformal invariance, which gets realized in the target space in the $\alpha^\prime\to \infty$ limit as higher spin symmetries. It is interesting to understand the manifestations of these gauge invariances. (d) These results seem to be robust: we find very parallel results in both Milne and the null orbifold.

\item The latter argument assumes that string amplitudes in flat space orbifolds are physically meaningful. Given the divergences in these amplitudes, one may object to this. Our standpoint in this paper is that these amplitudes are legitimate objects to study, but their breakdown indicates various physical phenomena. This is the same perspective adopted in \cite{HoroPolch} (and the numerous other papers which study these amplitudes) where these divergences were attributed to black hole production due to uncontrolled backreaction. 

\item One of our observations is that for large enough dimensionless $\alpha^\prime$, all the remaining amplitude divergences have a sensible IR interpretation. Since scattering amplitudes capture gauge invariant information, we expect its well behavedness at large $\alpha^\prime$ is meaningful in a higher spin interpretation. 

\item We emphasize that this is merely {\em one} paradigm for singularity resolution. For one, this approach is not immediately applicable to singularities in general relativity because the $\alpha^\prime\to \infty$ limit is the precise opposite limit of the classical GR limit in string theory. For another, there are other ways in which singularities can be resolved: namely when states become massless at points of the moduli space, the effective action will acquire singularities \cite{Strominger}. This means we have to add those degrees of freedom to the effective action to resolve the singularity. 

\item What classes of boundary conditions one should allow for the higher spin solutions we consider is a question worthy of exploration. The resolutions we consider do not fall within the asymptotically ``flat" boundary conditions of \cite{Arjun, Troncoso}. But after the first version of this paper appeared on the arXiv, it has been pointed out to us that both the singular geometry as well as our resolution do fall into a more general class of boundary conditions\footnote{We thank Daniel Grumiller for discussions and for sharing a draft of their upcoming work prior to publication.}. Happily, the gauge transformation that relates them within this class has zero canonical charges and is therefore trivial, i.e., the singular and resolved geometries are really the same state. This is precisely what one would expect since the string amplitude captures gauge-invariant information.


\item We will conclude with some comments about the zero of the higher spin field. There are two possible questions here. One is to see whether there exists {\em any} higher spin gauge transformation that can give rise to a resolved metric and a higher spin field without any zeros. We have stuck to the simplest resolution, so it is unclear if this is an artifact of the choices we made. It would be interesting (and within reach, one suspects) to have a concrete construction here, or a no-go theorem. Whatever the answer to this question is, it is unclear if there is a direct link between having a zero in the higher spin field and the existence of some pathology: keep in mind that regular metrics with everywhere zero higher spin fields, are considered non-singular configurations. It should also be kept in mind that vanishing of a field component is {\em not} necessarily a sign of a pathology (even though it can be): ingoing null coordinates have a zero at the horizon, but there is no singularity there; most metric components of Schwarzschild are identically zero, and yet the only point where it is singular is at $r=0$. In short, settling this question for higher spins is likely difficult given the absence of a higher spin version of Riemannian geometry.
\end{itemize}

Even though we were quite critical of our approach to higher spin singularity resolution in this section and our conclusions could be strengthened, we believe there is sufficient evidence at this point to indicate that the resolution is a real phenomenon and that this line of investigation deserves further exploration. In particular, it would be very interesting to understand what are the vestiges (if any) of this in the symmetry-broken phase of higher spin theory where the fields are massive, the gauge symmetry is hidden and the questions are more phenomenologically relevant. 

\acknowledgments
We thank Abhishake Sadhukan for collaboration at the early stages of this project. CK thanks Ben Craps, Justin David, Matthias Gaberdiel, Carlo Iazeolla, Alex Maloney and Joris Raeymaekers for discussions/correspondence, and Martin Schnabl for hospitality at IoP (Prague) during part of this work. The work of JS was partially supported by the Science and Technology Facilities Council (STFC) [grant number ST/J000329/1]. \\

\appendix

\section{Matrix representation}
\label{sec:matrix}

We provide an explicit representation of the $\fsl(3,\RR)$ matrices used in the higher spin calculations in section~\ref{sec:resolution}.
For a more detailed explanation of these conventions, see appendix A in \cite{Maloney}.

The generators of $\fsl(3,\RR)$ are denoted by $L_i$ and $W_a$, with $i=\pm1, 0$ and $a=\pm 2, \pm 1,0$. They satisfy
\begin{equation}
\begin{split}
  [L_i,\,L_j] &= (i-j)L_{i+j}\,, \\
  [L_i,\,W_a] &= (2i-a)W_{i+a}\,, \\
  [W_a,\,W_b] &= - \frac{1}{3}(a-b)(2a^2+2b^2-ab-8)L_{a+b}\,.
\end{split}
\end{equation}
Notice the set $\{L_{\pm 1},\,L_0\}$ generates $\fsl(2,\RR)$, whereas $\{W_a\}$ are the generators in $\fsl(3,\RR)$ not belonging to $\fsl(2,\RR)$.
The non-trivial components of the Lie algebra metric are
\begin{equation}
\begin{split}
  \text{tr}\left(L_0L_0\right)=2\,, \quad \quad &  \text{tr}\left(L_1L_{-1}\right)=-4\,, \\
  \text{tr}\left(W_0W_0\right)=\frac{8}{3}\,, \text{tr}\left(W_1W_{-1}\right)&=-4\,, \quad \text{tr}\left(W_2W_{-2}\right)=16\,.
\end{split}
\end{equation}
In section~\ref{sec:resolution}, we keep the generators $W_a$ and use the $\fsl(2,\RR)$ basis 
\begin{equation}
  T_0 = \frac{1}{2}\left(L_1 + L_{-1}\right)\,, \quad T_1 = \frac{1}{2}\left(L_1 - L_{-1}\right)\,, \quad T_2=L_0\,.
\end{equation}

An explicit representation of all these matrices is given below
\begin{equation}
\begin{split}
  L_1 = \begin{bmatrix}
  0 & 0 & 0 \\
  1 & 0 & 0 \\
  0 & 1 & 0
  \end{bmatrix}\,,
  L_0 &= \begin{bmatrix}
  1 & 0 & 0 \\
  0 & 0 & 0 \\
  0 & 0 & -1
  \end{bmatrix}\,,
  L_{-1} = \begin{bmatrix}
  0 & -2 & 0 \\
  0 & 0 & -2 \\
  0 & 0 & 0
  \end{bmatrix}\,, \\
  W_2 = 2\begin{bmatrix}
  0 & 0 & 0 \\
  0 & 0 & 0 \\
  1 & 0 & 0
  \end{bmatrix}\,,
  W_1 &= \begin{bmatrix}
  0 & 0 & 0 \\
  1 & 0 & 0 \\
  0 & -1 & 0
  \end{bmatrix}\,,
  W_{0} = \frac{2}{3} \begin{bmatrix}
  1 & 0 & 0 \\
  0 & -2 & 0 \\
  0 & 0 & 1
  \end{bmatrix}\,, \\
   \quad W_{-1} = \begin{bmatrix}
  0 & -2 & 0 \\
  0 & 0 & 2 \\
  0 & 0 & 0
  \end{bmatrix}\,, & \quad \quad
  W_{-2} = 2 \begin{bmatrix}
  0 & 0 & 4 \\
  0 & 0 & 0 \\
  0 & 0 & 0
  \end{bmatrix}\,.
\end{split}
\end{equation}

\section{Kundt Geometry and Vanishing Scalar Invariant (VSI) Spacetimes}
\label{sec:kundt}

To prove that the resolved metric \eqref{resmetr} has vanishing scalar curvature invariants constructed out of the Riemann tensor and its covariant derivatives, we first show that our metric falls into the Kundt family of spacetimes \cite{Podolsky} discussed in section (3.2) in \cite{Coley}. Our metric is of the form
\bea
ds^2=-du^2-2 du dr +\frac{dx^2}{f(u)^2}.
\eea
Defining $X=x/f(u)$ brings the metric to the form 
\bea
ds^2=-\left(1-\frac{X^2 f'^2}{f^2}\right)du^2-2 du dr + 2 \frac{X f'}{f} du dX + d X^2
\eea
which is of the form (25) in \cite{Coley} with $s=0$ and $\sigma=0$. This is precisely the Vanishing Scalar Invariant case of the Kundt metric as discussed in \cite{Coley}. It is easy to check that our metric is also a pp-wave, which is a special case of the Kundt geometry \cite{Podolsky}.

\section{Vertex Operators, OPEs and Infrared Divergences}
\label{sec:vertex}

In this appendix we will show that the $\alpha=0$ divergences discussed in subsection \ref{sec:alpha} are an IR effect. To do this, we closely follow the strategy in Appendix A and B of \cite{Ben}. The idea is to identify the divergent pieces in the amplitude integral by doing pairwise OPEs of the vertex operators in the 4-pt amplitude. 

The tachyon vertex operator in the null orbifold was first written in  \cite{LMS}  (again we refer the reader to \cite{LMS} for more details on the definitions and conventions):
\bea 
V_{p^+_i,J_i} (z) =  {1 \over \sqrt{2 \pi p_i^+}}\ \int_{-\infty}^\infty dp_i \,
 e^{-ip_i \xi_i} \, e^{i \vec{p_i} \cdot \vec{X}(z)}, \qquad \xi_i=-{J_i \over p^+_i} \label{vertex}
\eea
where
\bea
e^{i \vec{p}_i \cdot \vec{X}(z)} \equiv \exp \left[-i p^+_i x^- - i p^-_i x^+ + i p_i x + i \vec{p}_{\perp i} \cdot \vec{x}_{\perp} \right]. \label{vo}
\eea
Doing the integral explicitly and using $p_i^-=(p_i^2+m_i^2)/2p^+_i$, the result is
\bea
V_{p^+_i,J_i}  = &{1 \over \sqrt{ix^+}}\ \exp \left[-ip_i^+x^--i{m_i^2\over 2p_i^+} x^+ + i{ p_i^+
 \over 2 x^+}(x-\xi_i)^2+ i \vec{p}_{\perp i} \cdot \vec{x}_{\perp} \right] \label{expvert}
\eea
This is the form that we will find useful in showing that the extrema pole divergences in subsection \ref{extremapole} have their origins far away from the singularity. 

First we compute the leading behaviour of the two-point OPEs (see e.g., 
 \cite{Kiritsis}): 
\bea
V_{p^+_1,J_1}(z_1) V_{p^+_2,J_2}(z_2) \sim \nonumber \hspace{4in} \\ \frac{1}{2\pi\sqrt{p_1^+p_2^+}}\int_{-\infty}^{\infty}dp_1 \ dp_2 \ e^{-i(p_1 \xi_1+p_2 \xi_2) } 
|z_{12}|^{\alpha'\left(-p_1^+\frac{(p_2^2+m_2^2)}{2p^+_2}-p_2^+\frac{(p_1^2+m_1^2)}{2p^+_1}+p_1p_2 + \vec{p}_{\perp 1} \cdot \vec{p}_{\perp 2}\right)}
\times \hspace{1in} \label{OPE}  \\
\times \exp\left[-i(p_1^++p_2^+)x_--i\left(\frac{p_1^2+m_1^2}{2p_1^+}+\frac{p_2^2+m_2^2}{2p_2^+}\right)x^++i(p_1+p_2)x + i(\vec{p}_{\perp 1} +  \vec{p}_{\perp 2})\cdot \vec{x}_\perp \right]\hspace{0.25in} \nonumber
\eea
To bring this to a form that is easily compared with the $\alpha=0$ divergences, we first change integration variables from $(p_1, p_2)$ to $(U, p)$ defined by
\bea
\frac{p_1^2}{2 p_1^+}+U_1+\frac{p_2^2}{2 p_2^+}+U_2=\frac{p^2}{2 (p_1^++p_2^+)}+U, \label{changeofvar} \ \ 
p_1+p_2=p.
\eea 
The idea guiding this change is the interpretation of the last line in \eqref{OPE} as a vertex operator of the same general form as \eqref{vo}, defined at $z_2$, so that $V(z_1) V(z_2) \sim {\rm (stuff)} \times V(z_2)$. We have introduced the potential energy variables\footnote{Notice these variables were denoted as $V_i$ in \cite{LMS}.} $U_i=m_i^2/2 p_i^+$, and $U$ can be thought of as $m^2/2(p_1^++p_2^+)$ where $m$ is  the mass parameter of the new vertex operator. 

The inverse of these transformations are
\bea
p_1(p,U)= \frac{p \ p_1^+}{p_1^++p_2^+} \pm \sqrt{\frac{2 p_1^+p_2^+}{p_1^++p_2^+}(U-U_1-U_2)}\,, \label{eq:ins1} \\
p_2(p,U)= \frac{p \ p_2^+}{p_1^++p_2^+} \mp \sqrt{\frac{2 p_1^+p_2^+}{p_1^++p_2^+}(U-U_1-U_2)}\,, \label{eq:ins2}
\eea
where both signs are correlated. An important observation in what follows is that
\begin{equation}
U = U_1 + U_2 + \frac{\left(p_1p_2^+-p_2p_1^+\right)^2}{2p_1^+p_2^+\left(p_1^++p_2^+\right)} \ge U_1 + U_2. 
\end{equation}
In terms of the new variables, the OPE becomes
\bea
V_{p^+_1,J_1}(z_1) V_{p^+_2,J_2}(z_2) \sim \nonumber \hspace{4in} \\
\sum_\pm \frac{1}{2\pi}\int_{U_1+U_2}^{\infty} dU\  {\cal J}(p_1,p_2;U,p)\ |z_{12}|^{\alpha'\left(p_1^+ U_1+p_2^+ U_2 -(p_1^++p_2^+) U + \vec{p}_{\perp 1} \cdot \vec{p}_{\perp 2}\right)} \times \hspace{1.6in}  \label{OPEnewvar}\\ \times \int_{-\infty}^{\infty} dp\ \exp\left[-i(p_1^++p_2^+)x_--i\left(\frac{p^2}{2 (p_1^++p_2^+)}+U\right)x^++ip x\right] e^{-i\big(\xi_1 \ p_1(p,U)+\xi_2 \ p_2(p,U)\big) } \nonumber
\eea
where ${\cal J}(p_1,p_2;U,p)$ is the absolute value of the Jacobian for the change of variables (note that it turns out to be independent of $p$)
\begin{equation}
{\cal J}(p_1,p_2;U,p)=\left|\frac{\partial (p_1, p_2)}{\partial (U,p)}\right|=\sqrt{\frac{p_1^+p_2^+}{2(p_1^++p_2^+)
(U-U_1-U_2)}}\,,
\end{equation}
and the sum is over the two solutions in \eqref{eq:ins1} and \eqref{eq:ins2}.

We will show that the square root in the Jacobian is ultimately responsible for the $\alpha=0$ divergences, and that this non-analyticity in the square root is an IR effect. To see the latter, work with the vertex operators of the form in \eqref{expvert}. Far away from the singularity (large $x^+$), the relevant part of the zero mode integral when evaluating a 3-pt function of the form $\langle V^* V_1 V_2\rangle$ takes the form
\begin{equation}
\sim \int_{-\infty}^{\infty}\frac{dx^+}{(x^+)^{3/2}}\exp  \left[-i\left(\frac{m^2}{2p^+}-\frac{m_1^2}{2p_1^+}-\frac{m_2^2}{2p_2^+}\right) x^+ \right] \sim \sqrt{U-U_1-U_2} \label{nonal}
\end{equation}
The final $\sqrt{U-U_1-U_2}$ factor is easy to obtain by rescaling the integration variable in the LHS so that one is left with an overall factor of $\sqrt{U-U_1-U_2}$ times a purely numerical integral\footnote{Which can be related to a (generalized) Fresnel integral and can be evaluated in closed form, but since it is a purely numerical factor, we will not keep track of it.}. However, this result is true only when $U > U_1+U_2$: the integral is zero when $U < U_1+U_2$. To see this, consider the following manipulations: 
\bea
\int_{-\infty}^{\infty} e^{i a z}{dz \over z^{3/2}} =\int_0^{\infty}e^{i a z}{dz \over z^{3/2}}-\int_0^{-\infty}e^{i a z'}{dz' \over z'^{3/2}}= \int_0^{\infty} (e^{i a z}+i e^{-i a z})\frac{dz}{z^{3/2}}
\eea 
In the first equality, we have split the integral on the LHS into two pieces and renamed the dummy variable in the second piece as $z'$. In the second equality, we have made the variable redefinition $z'=e^{i \pi} \ z$ and appropriately changed the boundaries. The final integral can be written as 
\bea
(1+i)\int_0^\infty \frac{dz}{z^{3/2}}\big(\cos (a z) + \sin (a z) \big) 
\eea
This expression vanishes if $a > 0$, as seen from $
\int_0^\infty \frac{dz}{z^{3/2}} \sin (z) = - \int_0^\infty \frac{dz}{z^{3/2}} \cos (z) = \sqrt{2 \pi}$,
a fact that can be deduced from generalized Fresnel integrals \cite{wiki}. This shows that the expression on the LHS in (\ref{nonal}) vanishes when $U < U_1+U_2$.
Noting that we obtained the result (\ref{nonal}) by approximating the vertex operator (\ref{expvert}) by its far-from-singularity form, we conclude that the non-alayticity is an IR effect.

Now, we relate this to the $\alpha=0$ divergences. In section \ref{sec:alpha}, we noted that the $q \to 0$ limit gives rise to log divergences when $\alpha=0$, i.e., when $U_1+U_2=U_3+U_4$ in the 4-pt amplitude. To understand them as IR divergences we can adapt the argument of \cite{Ben} straightforwardly. First we note that if instead of the OPE considered so far, namely $V_1 V_2 \sim V$, we had looked at $V_1 V_2^* \sim V$ corresponding to an outgoing particle-2, we would have found $U \le U_1-U_2$ instead of $U \ge U_1 +U_2$ as the non-vanishing condition. To show this only requires to keep track of the signs in the exponents of the vertex operators. Notice the integration limits in the OPE \eqref{OPEnewvar} for $U$ also get appropriately modified when we consider the $V_1 V_2^* \sim V$ channel.

\begin{figure}[h]
\centering
\includegraphics[width=0.4\textwidth]{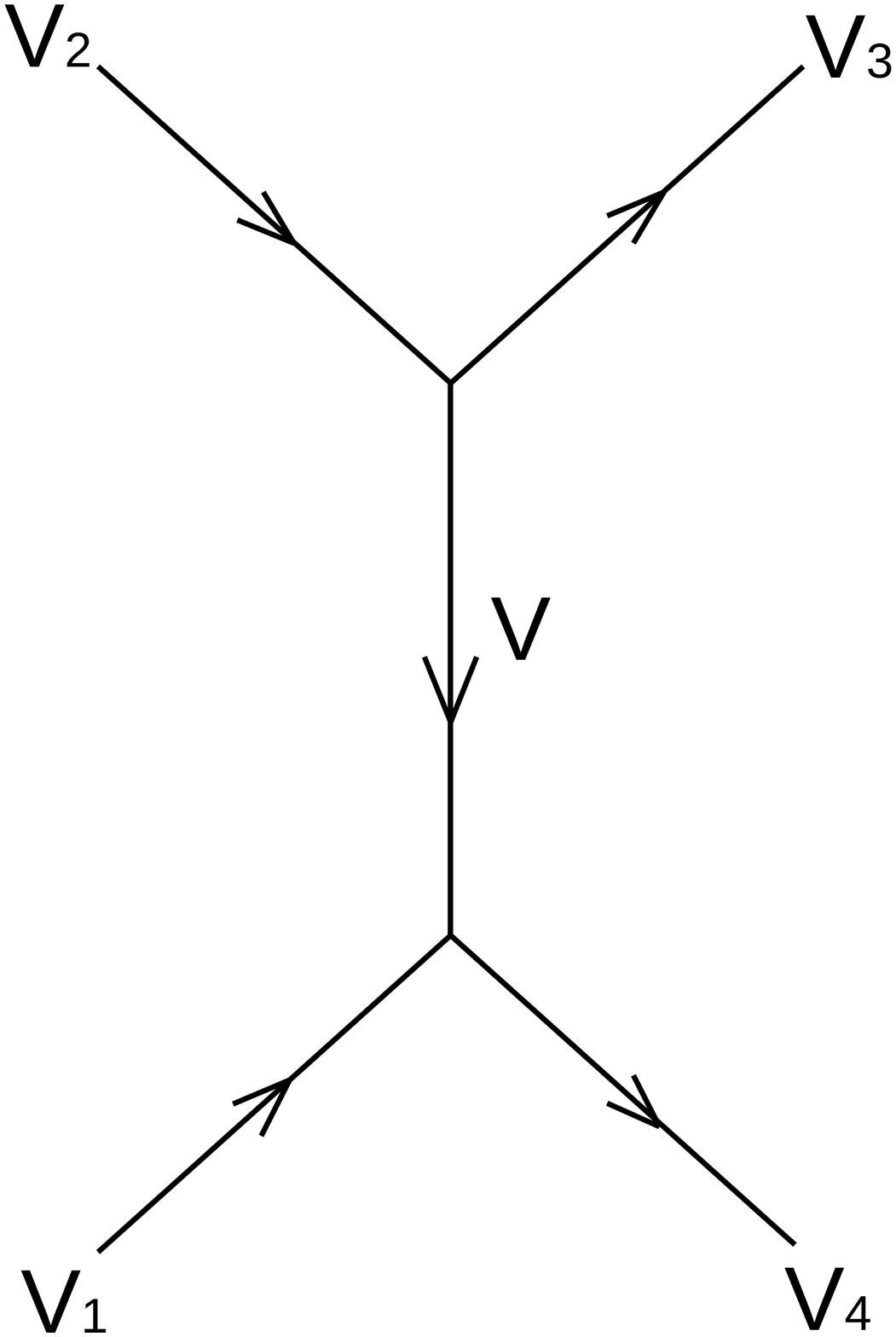}
\caption{Schematic diagram of the vertex operators in the 4-pt amplitude, relevant to the discussion about the $\alpha=0$ divergence in the Appendix.}
\label{ampfig}
\end{figure}
Now consider the schematic representation of our amplitude (not to be confused with field theory Feynman diagrams) in Figure \ref{ampfig}. Assume, for concreteness, that $U_2 \ge U_3$ and think of the process as $V_2$ turning into $V_3$ by emitting $V$, which in turn is absorbed by $V_1$, itself turning into $V_4$. From the schematic OPE for the $V_2 V_3^* \sim V$ part of the diagram, it follows that the potential energies must satisfy 
\begin{equation}
U \le U_2-U_3, \ \ {\rm or} \ \ -U \ge U_3 -U_2.
\end{equation}
Similarly, from the $V_1 V_4^* \sim V^*$ part of the diagram, we conclude that 
\begin{equation}
-U \le U_1 -U_4.
\end{equation}
Putting the two together, the non-vanishing condition for the amplitude is $U_3-U_2 \le -U \le U_1-U_4$, or $U_1+U_2 \ge U_3+U_4$. 

Now, in the 4-pt amplitude, the divergence around $\alpha=0$ for (otherwise) generic kinematics can be seen to arise from the Jacobian contributions in the OPE. From the $V_2 V_3^* \sim V$ part of the diagram, we see that the relevant piece of the integral is (we have kept track of the integration limit analogous to the situation in \eqref{OPEnewvar})
\begin{equation}
\int^{U_2-U_3} dU \frac{1}{\sqrt{U_2-U_3-U}} \times \frac{1}{\sqrt{U_4-U_1-U}}
\end{equation}
The second piece comes from the Jacobian for the $V_1 V_4^* \sim V^*$ part of the diagram. It is now straightforward to check that this integral can be divergent at the upper integration limit, only if $U_2-U_3=U_4-U_1$, which is precisely the condition equivalent to $\alpha=0$. Note that the integral can in principle receive further divergences in the $\alpha=0$ case (from other regions of the integral) when the parameters are further tuned, as we saw in the discussion about the boundary pole divergences.

\bibliographystyle{JHEP}
\bibliography{LMS-Joan}

\end{document}